\documentstyle[preprint]{aastex}

\def  \be	 {\begin{equation}}
\def  \ee	 {\end{equation}}
\def  \beq	 {\begin{eqnarray}}
\def  \eeq	 {\end{eqnarray}}
\def  \simlt	 {{\lower.5ex\hbox{$\;\buildrel{\mbox{\footnotesize$<$}}\over{\mbox{\footnotesize$\sim$}}\;$}}}
\def  \simgt	 {{\lower.5ex\hbox{$\;\buildrel{\mbox{\footnotesize$>$}}\over{\mbox{\footnotesize$\sim$}}\;$}}}
\def  \define	 {\stackrel{\rm def}{=}}
\def  \bfdelta	 {{\mbox{\boldmath$\delta$}}}

\begin{document}

\title{\bf On the thermal conduction in tangled magnetic fields in clusters of galaxies}

\author{Leonid~Malyshkin}

\affil{Princeton University Observatory, Princeton NJ 08544, USA}

\email{leonmal@astro.princeton.edu}

\date{\today}

\begin{abstract}
Thermal conduction in tangled magnetic fields is reduced because heat conducting electrons must travel 
along the field lines longer distances between hot and cold regions of space than if there were no fields. 
We consider the case when the tangled magnetic field has a weak homogeneous component. We examine two 
simple models for temperature in clusters of galaxies: a time-independent model and a time-dependent one. 
We find that the actual value of the effective thermal conductivity in tangled magnetic fields depends on 
how it is defined for a particular astrophysical problem. Our final conclusion is that the heat conduction 
never totally suppressed but is usually important in the central regions of galaxy clusters, and therefore, 
it should not be neglected.
\end{abstract}


%************************************************************************************************************
%************************************************************************************************************
%************************************************************************************************************

\section{Introduction}\label{INTRODUCTION}

In order to solve the problem of electron thermal conduction in a stochastic magnetic field, one should 
consider separately two effects that reduce the conduction (Pistinner \& Shaviv 1996; Chandran \& Cowley 1998).
The first effect is that the heat conducting electrons have to travel along tangled magnetic 
field lines, and therefore, they have to travel longer distances between hot and cold regions of 
space (Tribble 1989, Tao 1995). The second effect is that electrons, while they are traveling along 
the field lines, become trapped and detrapped between magnetic mirrors, regions of strong magnetic field 
(Chandran, Cowley, \& Ivanushkina 1999). 

In our recent paper (Malyshkin \& Kulsrud 2000) we studied the second effect, and we found the reduction of
the thermal conductivity parallel to the magnetic field lines, $\kappa_\parallel$, relative to the Spitzer 
value for the thermal conductivity, $\kappa_{\rm S}$, caused by the presence of magnetic mirrors. 
In this paper we consider the first effect, and we calculate the further reduction of thermal conduction 
relative to $\kappa_\parallel$, caused by the tangled structure of the magnetic field lines. As a result, 
we obtain the total reduction caused by the both effects, and we calculate the effective thermal conductivity 
$\kappa_{\rm eff}$, which one has to apply for astrophysical problems. It turns out that there is no an 
universal definition of $\kappa_{\rm eff}$, and the result depends on the particular astrophysical problem 
under consideration.

The structure of this paper is the following. In the next section we consider the random walk of tangled 
magnetic field lines by making use of a diffusion approximation for this random walk. We find the 
expressions for the probability distributions of the field line lengths and for the Laplace transform, in
the field line length, of these distributions. In Section~\ref{MODEL} we use the Laplace transforms to 
calculate the effective thermal conductivity for our stationary model of a galaxy cluster. In 
Section~\ref{T_EVOLUTION} we consider a time-dependent model and study the time evolution of temperature 
in a galaxy cluster. Finally, in Section~\ref{DISCUSSION} we discuss our results, compare them with the 
results reported in previous papers, and consider the applications to cooling flow problem. For two 
opposite limiting cases our results for $\kappa_{\rm eff}$ are: one is similar to that of Tribble (1989),
and the other coincides to that of Tao (1995).

%************************************************************************************************************
%************************************************************************************************************
%************************************************************************************************************

\section{Diffusion approximation for the random walk of magnetic field lines}\label{RANDOM_WALK}

In a particularly simple approach, the behavior of tangled magnetic field lines can be considered by using the 
one-dimensional random walk model suggested by Tribble (1989). In this model the lines are assumed to random 
walk between two infinite boundary plates, which are placed at $x=0$ and $x=X_0$ perpendicular to the 
$x$-direction. In this section we consider this simple model in order to find the probability distributions of 
the lengths of the field lines in clusters of galaxies (we then use these distributions to find the effective 
thermal conductivity in the next sections). Contrary to the discrete calculations of Tribble, we use the 
continuum diffusion approximation for the random walk of the field lines. This allows us to include a weak 
homogeneous mean magnetic field component into our calculations.

%************************************************************************************************************
\subsection{Diffusion equation}\label{DIFFUSION_EQUATION}

We assume that the mean magnetic field component, $\bf\langle B\rangle$, is homogeneous, 
i.e.~${\bf\langle B\rangle}={\rm const}$. Let choose the coordinate system in a such way that 
$\bf\langle B\rangle$ lies in the $x$-$z$-plane: ${\langle B\rangle}_x=\langle B\rangle\cos\theta$, 
${\langle B\rangle}_y=0$, ${\langle B\rangle}_z=\langle B\rangle\sin\theta$. Here $\theta$
is the angle between the mean field component and the $x$-direction. 
Further, assume that the random component of the magnetic field, $\bfdelta\bf B$, is much stronger 
than the mean component, i.e.~$\langle B\rangle\big/\delta B\ll 1$. Let express $x$-, $y$- and $z$-components 
of $\bfdelta\bf B$ in the spherical system of coordinates: ${\delta B}_x=\delta B\cos\varphi$, 
${\delta B}_y=\delta B\sin\varphi\cos\psi$ and ${\delta B}_z=\delta B\sin\varphi\sin\psi$. 
The random component is isotropically distributed, so
$\cos\varphi$ and $\psi$ are uniformly distributed over $[-1,1]$ and $[0,2\pi)$ respectively.
Under these assumptions, the cosine of the angle between the total magnetic field, 
${\bf B}={\bfdelta\bf B}+{\bf\langle B\rangle}$, and the $x$-direction is
\beq
\cos\alpha=\cos\varphi+\bigl[\langle B\rangle\big/\delta B\bigr]
\bigl[\cos\theta-\cos\theta\cos^2\!\varphi-\sin\theta\sin\varphi\cos\varphi\sin\psi\bigr],
\label{COS_ALPHA}
\eeq
where we keep terms only up to first order in $\langle B\rangle\big/\delta B\ll 1$.

Let the largest scale of the random component of the magnetic field be $l_0\ll X_0$. The largest scale component 
has more magnetic energy in it than all smaller scale components have. Therefore, the decorrelation length of 
the total random component is $l_0$, and over each decorrelation length the random field changes into an 
entirely new direction. As a result, the mean step and the mean-squared step of the field line 
random walk in the $x$-direction 
are
\beq
\begin{array}{rclcl}
\langle\Delta x\rangle &=& l_0{\langle\cos\alpha\rangle}_{\varphi,\psi,\,\delta B} &=& (2/3)l_0\epsilon\cos\theta,
\\
\langle(\Delta x)^2\rangle &=& l_0^2{\langle\cos^2\!\alpha\rangle}_{\varphi,\psi,\,\delta B} &=& (1/3)l_0^2
\end{array}
\label{DELTA_X}
\eeq
over each decorrelation length.
Here, to obtain these final expressions, we average $\cos\alpha$ and $\cos^2\!\alpha$ over direction and 
over absolute value of the field random component (i.e.~over $\varphi$, $\psi$ and $\delta B$). 
We also introduce the parameter $\epsilon\define\langle B\rangle\big/\langle\delta B\rangle\ll 1$,
which is the mean ratio of the magnetic field mean component to the random component.

\begin{figure}
\vspace{7.0cm}
\includegraphics{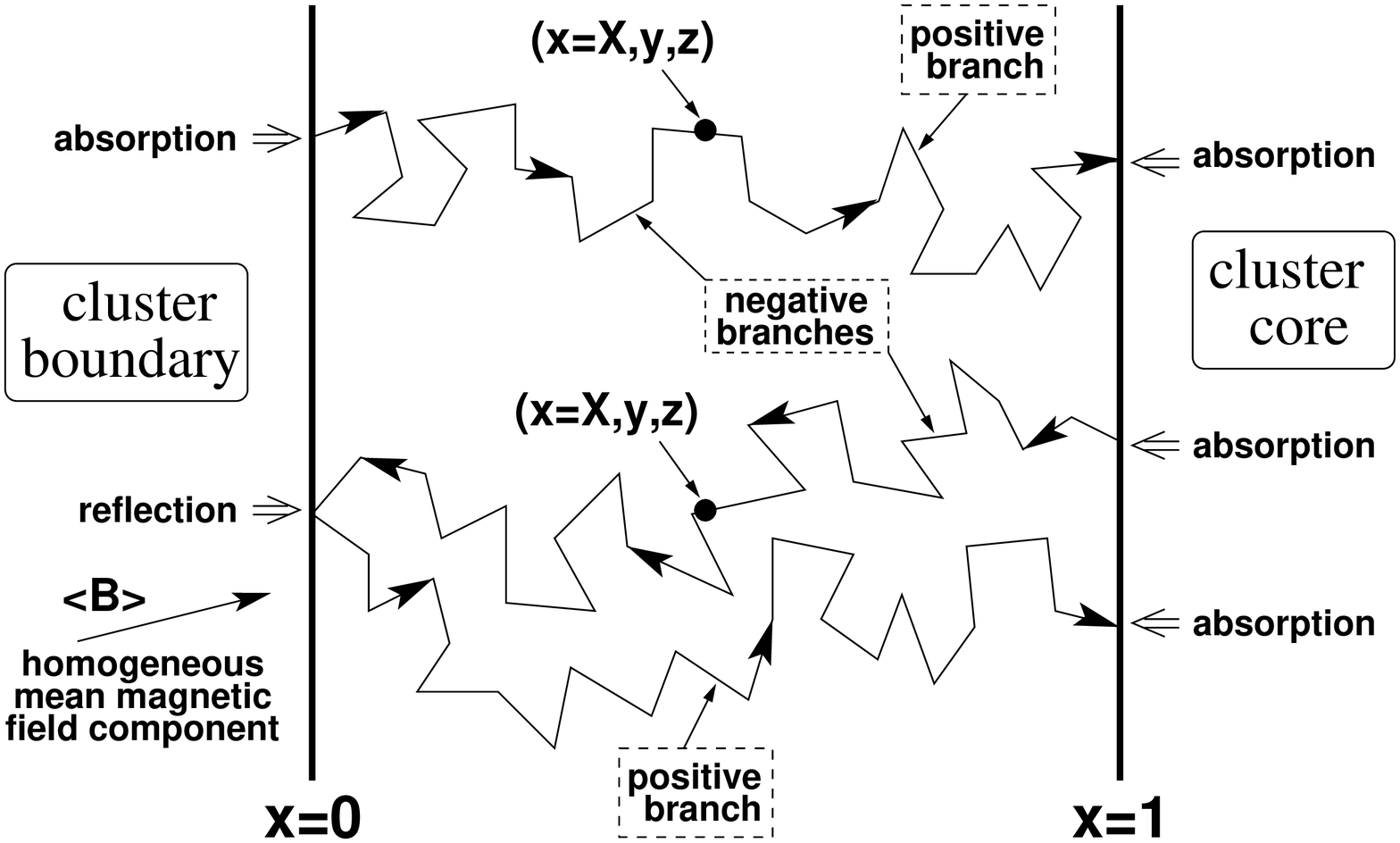}
\caption{There is the unique magnetic field line that goes through any given point $({\sf x}={\sf X},y,z)$ of 
space, this field line leaves the point along the positive branch and along the negative branch. 
All positive branches are reflected at the cluster boundary at ${\sf x}=0$, and are absorbed at the cluster 
core at ${\sf x}=1$. All negative branches are absorbed both at the boundary and at the core. 
Thus, there are two types of magnetic field lines: first, there are lines that go from the cluster boundary to 
the cluster core (see the upper line in this figure), and second, there are lines that leave the core and 
finally come back to the core (see the lower line in this figure).}
\label{FIG_PICTURE}
\end{figure}

There is the unique magnetic field line that goes through any given point of space between the two boundary 
plates. This field line leaves the point along two branches: a {\it positive branch} that starts at the initial 
point and goes always in the direction of the local field, and a {\it negative branch} that goes always opposite 
to the direction of the local field (see Figure~\ref{FIG_PICTURE}). Let consider the following problem. 
Start at point $(x=X,y,z)$ and follow the positive branch of the magnetic field line going through this point, 
i.e.~along the field line and always in the direction of the local field. 
The positive field line branch random walks in space according to equations~(\ref{DELTA_X}). 
Let $P^+(s,x)\,ds\,dx$ be the probability that we are at $x$-position $x\in[x,x+dx)$ [and at
any $y$-, $z$-positions] after we ``have walked'' along the line a distance~$s\in[s,s+ds)$. 
The upper index ``$(+)$'' refers to the positive branch of the magnetic field line.
Using equations~(\ref{DELTA_X}), it is straightforward to write the diffusion equation for $P^+(s,x)$
as~\footnote{
Note that the diffusion approximation is valid if $x\gg l_0$ and $X_0-x\gg l_0$.
}
\beq
\frac{\partial P^+}{\partial s} &=& -\frac{\langle\Delta x\rangle}{l_0}\frac{\partial P^+}{\partial x}+
\frac{1}{2}\frac{\langle(\Delta x)^2\rangle}{l_0}\frac{\partial^2 P^+}{\partial x^2}
=-\frac{2\epsilon\cos\theta}{3}\frac{\partial P^+}{\partial x}+\frac{l_0}{6}\frac{\partial^2 P^+}{\partial x^2},
\label{DIFFUSION_EQ_DIMENTIONAL}\\
P^+(0,x) &=& \delta(x-X).
\label{INIT_CONDITION_DIMENTIONAL}
\eeq
This is a Fokker-Planck equation.
Here, $\delta(x-X)$ is the Dirac delta-function, and the formula on the second line is the initial 
condition on $P^+(s,x)$, which means we start walking along the field line at $x=X$.
If we now consider the field line's negative branch, which goes always opposite to the direction 
of local field, then we need to replace $\epsilon$ by $-\epsilon$ in formula~(\ref{DIFFUSION_EQ_DIMENTIONAL})
in order to obtain the differential equation for the probability function $P^-(s,x)$ for this negative 
branch. (Note, that this replacement corresponds to the substitution $\pi-\theta$ for $\theta$.) 

Hereafter, we use convenient dimensionless variables 
\beq
{\sf s}={(3X_0^2/l_0)}^{-1}s, \quad {\sf x}=x/X_0, \quad {\sf X}=X/X_0, \quad  {\sf P}^\pm=X_0 P^\pm.
\label{VARIABLES}
\eeq
Note that the positions of the boundary plates are ${\sf x}=0$ and ${\sf x}=1$. In these variables
\beq
\frac{\partial {\sf P}^\pm}{\partial {\sf s}} = \mp\beta\frac{\partial {\sf P}^\pm}{\partial {\sf x}}+
\frac{1}{2}\frac{\partial^2 {\sf P}^\pm}{\partial {\sf x}^2}, & 
\qquad & {\sf P}^\pm(0,{\sf x}) = \delta({\sf x}-{\sf X}).
\label{DIFFUSION_EQ}
\eeq
Here, we introduce an important parameter
\beq
\beta\define 2X_0l_0^{-1}\epsilon\cos\theta.
\label{BETA}
\eeq

In the limit $\beta\ll 1$, the effect of the mean magnetic field component is negligible and the ``diffusion term'' 
$(1/2)\,(\partial^2 {\sf P}^\pm/\partial {\sf x}^2)$ in equation~(\ref{DIFFUSION_EQ}) is dominant. In this
small $\beta$ limit the random walk of field lines is controlled by their diffusion in space, and
for a given ${\sf s}\sim 1$, the functions ${\sf P}^\pm({\sf s},{\sf x})$ have considerable spreads in ${\sf x}$. 
On the other hand, in the limit $\beta\gg 1$, the effect of the mean magnetic field component is large and the 
``flux term'' $-\beta\,(\partial {\sf P}^\pm/\partial {\sf x})$ in equation~(\ref{DIFFUSION_EQ}) is dominant. In 
this large $\beta$ limit the probability functions ${\sf P}^\pm$ stay very narrow in $x$ for a given 
value of $\sf s$, ${\sf P}^\pm({\sf s},{\sf x})\approx \delta({\sf x}-{\sf X}\mp\beta {\sf s})$.
When $\beta\sim 1$, both terms in the right-hand side of equation~(\ref{DIFFUSION_EQ}) are equally 
important. Thus, $\beta$ indicates whether the influence of the mean field component on the field line 
random walk is negligible ($\beta\ll 1$), important ($\beta\sim 1$) or dominant ($\beta\gg 1$).

%************************************************************************************************************
\subsection{Boundary conditions and probability fluxes}\label{BOUNDARY}

Equations~(\ref{DIFFUSION_EQ}) need the boundary 
conditions on the probability functions ${\sf P}^\pm({\sf s},{\sf x})$ at the two boundary plates. Let assume 
for definiteness, that the plate at ${\sf x}=0$ represents the boundary of a galaxy cluster, the plate at 
${\sf x}=1$ represents the cluster core (see Figure~\ref{FIG_PICTURE}), 
and the mean component of the magnetic field has a positive $x$-component, i.e.~$0<\theta<\pi/2$ (clearly the 
sign of $\cos\theta$ has no bearing on thermal diffusion problem). 
As we said above, there are two field line branches that we can follow starting at ${\sf x}={\sf X}$, the positive 
branch and the negative branch. Each branch random walks in space, and can reach the cluster boundary at 
${\sf x}=0$ and the cluster core at ${\sf x}=1$.

Let first consider the boundary conditions at the cluster core. The core is assumed to be much denser and much 
colder than the rest of the cluster, so we can neglect the temperature $T$ in the core and set it to zero. 
As a result, whenever the positive or the negative branch reaches the core at ${\sf x}=1$, it looses its thermal
energy and cools down to zero temperature immediately. 
Therefore, there is no any point in following the field line branches after they first reach the core, and the 
corresponding boundary conditions on the probability functions ${\sf P}^\pm$ at the core are the absorption 
conditions: ${\sf P}^\pm({\sf s},1)=0$.

Let now consider the boundary conditions at the cluster boundary. We assume that the density drops significantly 
there. Therefore, the random component of the magnetic field also drops considerably at the 
cluster boundary, because this random component is believed to be created by MHD dynamo action inside the cluster.
On the other hand, the mean component of the field hardly changes at the boundary, because it is associated 
with the open magnetic field lines that leave the cluster. As a result, the parameter
$\beta$, given by equation~(\ref{BETA}), increases and becomes large at the cluster boundary
(while the mean field component can still be less than the random component, so that our diffusion
approximation is still valid). In other words, we consider $\beta$ to be constant inside the cluster and to 
become large at the boundary. Therefore, at the cluster boundary ${\sf x}=0$, the term 
$-\beta\,(\partial {\sf P}^\pm/\partial {\sf x})$ in equation~(\ref{DIFFUSION_EQ}) is always dominant.
As a result, the positive field line branch is reflected at the cluster boundary,
while the negative branch is absorbed and leaves the cluster.

To summarize, we write the boundary conditions for the positive and negative field line branches as
\beq
\begin{array}{rl}
{\sf P}^-({\sf s},0)=0, & \mbox{absorption of the negative branch at the boundary},\\
{\sf P}^-({\sf s},1)=0, & \mbox{absorption of the negative branch at the core};
\end{array}
\label{BOUNDARY_MINUS}
\eeq
\beq
\begin{array}{rl}
\beta\, {\sf P}^+({\sf s},0)-(1/2){(\partial {\sf P}^+/\partial {\sf x})|}_{{\sf x}=0}=0, & 
\mbox{reflection of the positive branch at the boundary},\\
{\sf P}^+({\sf s},1)=0, & \mbox{absorption of the positive branch at the core},
\end{array}
\label{BOUNDARY_PLUS}
\eeq
see Figure~\ref{FIG_PICTURE}.

Let for a moment consider the positive field line branch and integrate the diffusion equation~(\ref{DIFFUSION_EQ}) 
over ${\sf x}\in(0,1)$, using boundary conditions~(\ref{BOUNDARY_PLUS}). As a result, we obtain the 
negative change of the total probability that we are still inside the cluster, ${\sf x}\in(0,1)$, 
after we walked a distance ${\sf s}$ along the line's positive branch:
\beq
-\frac{d}{d{\sf s}}\int_0^1 {\sf P}^+({\sf s},{\sf x})\,d{\sf x}=
-{\left.\frac{1}{2}\frac{\partial {\sf P}^+}{\partial {\sf x}}\right|}_{{\sf x}=1}
={\sf F}_1^+({\sf s}).
\label{PROBABILITY_CHANGE_PLUS}
\eeq
Here we introduce a ``probability flux'' into the cluster core at ${\sf x}=1$:
\beq
{\sf F}_1^+({\sf s})=-(1/2)\,{(\partial {\sf P}^+/\partial {\sf x})|}_{{\sf x}=1}
\label{F_PLUS}
\eeq
(the upper index ``$+$'' refers to the field line's positive branch). 
Note, that the probability flux across the cluster boundary at ${\sf x}=0$
is zero because of the reflection condition there [see eqs.~(\ref{BOUNDARY_PLUS})].
The probability, that we leave the cluster and enter the cluster core when ${\sf s}\in[{\sf s},{\sf s}+d{\sf s})$, 
is equal to ${\sf F}_1^+({\sf s})\,d{\sf s}$. Therefore, ${\sf F}_1^+({\sf s})$ is the 
{\it probability distribution} of the lengths $\sf s$ of all positive field line branches that start 
at ${\sf x}={\sf X}$, random walk in space and finally come to the cluster core at ${\sf x}=1$. 

Now consider the negative field line branch. In this case, we integrate the appropriate diffusion 
equation~(\ref{DIFFUSION_EQ}) over ${\sf x}\in(0,1)$ and use the boundary conditions~(\ref{BOUNDARY_MINUS}).
We find that the negative change of the total probability is now given by the sum of two probability 
fluxes ${\sf F}_1^-({\sf s})$ and ${\sf F}_0^-({\sf s})$ into the core (at ${\sf x}=1$) and across 
the boundary (at ${\sf x}=0$) respectively:
\beq
{\sf F}_1^-({\sf s})=-(1/2)\,{(\partial {\sf P}^-/\partial {\sf x})|}_{{\sf x}=1},
\qquad
{\sf F}_0^-({\sf s})=(1/2)\,{(\partial {\sf P}^-/\partial {\sf x})|}_{{\sf x}=0}.
\label{F_MINUS}
\eeq
Similarly to the case of the positive branch, 
${\sf F}_1^-({\sf s})$ and ${\sf F}_0^-({\sf s})$ are the probability distributions of the 
lengths of all negative field line branches that start at ${\sf x}={\sf X}$, random walk in space and finally 
come to either the cluster boundary at ${\sf x}=0$, or to the cluster core at ${\sf x}=1$. 

Because the probability fluxes are the probability distributions of the lengths of 
field line branches, hereafter, we refer to them as to the probability distributions.

%************************************************************************************************************
\subsection{Laplace transform solutions}

Although equations~(\ref{DIFFUSION_EQ}) are linear, they are still difficult to solve analytically for
both sets of boundary conditions given by equations~(\ref{BOUNDARY_MINUS}) and~(\ref{BOUNDARY_PLUS}).
We solve equations~(\ref{DIFFUSION_EQ}) numerically in Section~\ref{T_EVOLUTION}. However, in some 
cases we do not need the solution of equations~(\ref{DIFFUSION_EQ}) in order to find the effective 
thermal conductivity in tangled magnetic fields. In the next section we will find this conductivity using a 
simple stationary model and the Laplace images of the probability distributions ${\sf F}_1^+({\sf s})$, 
${\sf F}_1^-({\sf s})$ and ${\sf F}_0^-({\sf s})$ that we obtain in this section.

To calculate these Laplace images, we first take the Laplace transforms ${\sf s}\to{\tilde{\sf s}}$, 
${\sf P}^\pm({\sf s},{\sf x})\to {\tilde{\sf P}}^\pm({\tilde{\sf s}},{\sf x})$ of equations~(\ref{DIFFUSION_EQ}). 
We have
\beq
{\tilde{\sf s}}{\tilde{\sf P}}^\pm-\delta({\sf x}-{\sf X}) &=& 
\mp\beta\frac{\partial {\tilde{\sf P}}^\pm}{\partial {\sf x}}+
\frac{1}{2}\frac{\partial^2 {\tilde{\sf P}}^\pm}{\partial {\sf x}^2}.
\label{L_DIFFUSION_EQ}
\eeq
The Laplace images ${\tilde{\sf P}}^\pm({\tilde{\sf s}},{\sf x})$ must be continuous functions of ${\sf x}$. Thus, 
integrating equations~(\ref{L_DIFFUSION_EQ}) across ${\sf x}={\sf X}$, i.e.~over an infinitesimal interval 
${\sf x}\in ({\sf X}-0,{\sf X}+0)$, we obtain an additional jump conditions together with the continuity conditions
\beq
\bigl[\partial{\tilde{\sf P}}^\pm/\partial {\sf x}\bigr]_{{\sf X}\pm 0} = -2, 
\qquad
\bigl[{\tilde{\sf P}}^\pm \bigr]_{{\sf X}\pm 0} = 0.
\label{L_JUMP_CONDITIONS}
\eeq
Here, we denote the jump at ${\sf x}={\sf X}$ as $[...]_{{\sf X}\pm 0}$.

The boundary conditions on the images ${\tilde{\sf P}}^\pm({\tilde{\sf s}},{\sf x})$ are the same as those 
on the original functions ${\sf P}^\pm({\sf s},{\sf x})$. In the case of the negative branch they are given by 
equations~(\ref{BOUNDARY_MINUS}), in the case of the positive branch they are given by 
equations~(\ref{BOUNDARY_PLUS}).
Equation~(\ref{L_DIFFUSION_EQ}) is a simple linear homogeneous differential equation provided ${\sf x}\ne {\sf X}$.
There are two independent solutions of this equation expressed in terms of hyperbolic functions.
We express the two general solutions of this equation in the regions $0\le {\sf x}<{\sf X}$ and ${\sf X}<{\sf x}\le 1$ 
as two linear combinations of these hyperbolic  solutions.
Then, we use the appropriate boundary conditions and jump conditions~(\ref{L_JUMP_CONDITIONS}) 
to find the unique continuous solutions of equations~(\ref{L_DIFFUSION_EQ}) for ${\sf P}^\pm({\sf s},{\sf x})$
in the whole interval of ${\sf x}$, $0\le {\sf x}\le 1$. We have
\beq
{\tilde{\sf P}}^+({\tilde{\sf s}},{\sf x})=
\frac{2}{\xi_0}\frac{\exp{[\beta({\sf x}-{\sf X})]}}{\xi_0\cosh\xi_0+\beta\sinh\xi_0}
\times\cases{
[\xi_0\cosh{(\xi_0{\sf X})}+\beta\sinh{(\xi_0{\sf X})}]\sinh{[\xi_0(1-{\sf x})]}, & $\!\! {\sf X}\le {\sf x}$; \cr
[\xi_0\cosh{(\xi_0{\sf x})}+\beta\sinh{(\xi_0{\sf x})}]\sinh{[\xi_0(1-{\sf X})]}, & $\!\! {\sf x}\le {\sf X}$ \cr
}
\label{L_P_PROB_PLUS}
\eeq
for the positive field line branch, and
\beq
{\tilde{\sf P}}^-({\tilde{\sf s}},{\sf x})=\frac{2}{\xi_0}\frac{\exp{[-\beta({\sf x}-{\sf X})]}}{\sinh\xi_0}
\times\cases{
\sinh{(\xi_0{\sf X})}\sinh{[\xi_0(1-{\sf x})]}, & ${\sf X}\le {\sf x}$; \cr
\sinh{(\xi_0{\sf x})}\sinh{[\xi_0(1-{\sf X})]}, & ${\sf x}\le {\sf X}$ \cr
}
\label{L_P_PROB_MINUS}
\eeq
for the negative branch. Here, we introduce
\beq
\xi_0({\tilde{\sf s}})\define \sqrt{\beta^2+2{\tilde{\sf s}}\,}.
\label{XI}
\eeq

Now, we substitute these formulas into Laplace transformed equations~(\ref{F_PLUS}) and~(\ref{F_MINUS})
to find the Laplace images ${\tilde{\sf F}}_1^+({\tilde{\sf s}})$, ${\tilde{\sf F}}_1^-({\tilde{\sf s}})$
and ${\tilde{\sf F}}_0^-({\tilde{\sf s}})$
of the probability distributions ${\sf F}_1^+({\sf s})$, ${\sf F}_1^-({\sf s})$ 
and ${\sf F}_0^-({\sf s})$. We have
\beq
{\tilde{\sf F}}_1^+({\tilde{\sf s}}) =
\int_0^\infty\!\! e^{-{\tilde{\sf s}}{\sf s}}\, {\sf F}_1^+({\sf s})\,d{\sf s}
\!\!&=&\!\! -\frac{1}{2}{\frac{\partial{\tilde{\sf P}}^+}{\partial {\sf x}}\Big|}_{{\sf x}=1}
=\frac{\xi_0\cosh{(\xi_0{\sf X})}+\beta\sinh{(\xi_0{\sf X})}}{\xi_0\cosh\xi_0+\beta\sinh\xi_0}\,e^{\beta(1-{\sf X})},
\label{L_F_PLUS_1}
\\
{\tilde{\sf F}}_1^-({\tilde{\sf s}}) = 
\int_0^\infty\!\! e^{-{\tilde{\sf s}}{\sf s}}\, {\sf F}_1^-({\sf s})\,d{\sf s} 
\!\!&=&\!\! -\frac{1}{2}{\frac{\partial {\tilde{\sf P}}^-}{\partial {\sf x}}\Big|}_{{\sf x}=1}
=\frac{\sinh{(\xi_0{\sf X})}}{\sinh\xi_0}\,e^{-\beta(1-{\sf X})},
\label{L_F_MINUS_1}
\\
{\tilde{\sf F}}_0^-({\tilde{\sf s}}) =
\int_0^\infty\!\! e^{-{\tilde{\sf s}}{\sf s}}\, {\sf F}_0^-({\sf s})\,d{\sf s} 
\!\!&=&\!\! +\frac{1}{2}{\frac{\partial {\tilde{\sf P}}^-}{\partial {\sf x}}\Big|}_{{\sf x}=0}
=\frac{\sinh{[\xi_0(1-{\sf X})]}}{\sinh\xi_0}\,e^{\beta {\sf X}}.
\label{L_F_MINUS_0}
\eeq
Here, the integrals are the definition of the Laplace images, and $\xi_0$ is given by formula~(\ref{XI}).

As is well known from the theory of the Laplace transform [and can easily be checked
by differentiation of equations~(\ref{L_F_PLUS_1})--(\ref{L_F_MINUS_0}) with 
respect to the Laplace variable $\tilde {\sf s}$ and setting $\tilde {\sf s}$ to zero], the following formulas 
stand for the integral moments of the probability distributions ${\sf F}_1^+({\sf s})$, ${\sf F}_1^-({\sf s})$ 
and ${\sf F}_0^-({\sf s})$:
\beq
\int_0^\infty {\sf s}^n {\sf F}_1^\pm({\sf s})\,d{\sf s}={(-1)}^n\,
\frac{d^n {\tilde{\sf F}}_1^\pm}{d{\tilde{\sf s}}^n}\Big|_{{\tilde{\sf s}}=0},
\qquad
\int_0^\infty {\sf s}^n {\sf F}_0^-({\sf s})\,d{\sf s}={(-1)}^n\,
\frac{d^n {\tilde{\sf F}}_0^-}{d{\tilde{\sf s}}^n}\Big|_{{\tilde{\sf s}}=0}.
\label{MOMENTS}
\eeq
For example, the zeroth integral moment of ${\sf F}_1^+({\sf s})$ is 
\beq
\int_0^\infty {\sf F}_1^+({\sf s})\,d{\sf s}={\tilde{\sf F}}_1^+(0)=1,
\label{FIRST_MOMENT_F_PLUS}
\eeq
which means that the total probability for the positive branch to ultimately reach the cluster core is unity,
as it should be because positive branches always end up in the core. The zeroth
integral moments of ${\sf F}_1^-({\sf s})$ and ${\sf F}_0^-({\sf s})$ are
\beq
\int\limits_0^\infty {\sf F}_1^-({\sf s})\,d{\sf s}={\tilde{\sf F}}_1^-(0)=
\frac{\exp{(2\beta {\sf X})}-1}{\exp{(2\beta)}-1},
\quad
\int\limits_0^\infty {\sf F}_0^-({\sf s})\,d{\sf s}={\tilde{\sf F}}_0^-(0)=
\frac{\exp{(2\beta)}-\exp{(2\beta {\sf X})}}{\exp{(2\beta)}-1},
\label{FIRST_MOMENT_F_MINUS}
\eeq
which means that ${\tilde{\sf F}}_1^-(0)$ and ${\tilde{\sf F}}_0^-(0)$ are the probabilities
for the negative branch to reach the cluster core (at ${\sf x}=1$) and the cluster boundary (at ${\sf x}=0$)
respectively. Note, that the total probability for the negative branch to reach either the core or the 
boundary, is equal to one, ${\tilde{\sf F}}_1^-(0)+{\tilde{\sf F}}_0^-(0)=1$, as it should be.

It follows from equations~(\ref{FIRST_MOMENT_F_MINUS}) that ${\sf F}_1^-({\sf s})/{\tilde{\sf F}}_1^-(0)$
and ${\sf F}_0^-({\sf s})/{\tilde{\sf F}}_0^-(0)$ are the {\it normalized} (to one) probability
distributions of the lengths of only those negative field line branches that reach the cluster core and
only those negative branches that reach the cluster boundary respectively (see Section~\ref{BOUNDARY}). 
The probability distributions of the positive branches, ${\sf F}_1^+({\sf s})$, does not need to be 
normalized, because all positive branches always end up in the cluster core.

%************************************************************************************************************
%************************************************************************************************************
%************************************************************************************************************

\section{Stationary model}\label{MODEL}

In this section we consider a stationary one-dimensional temperature distribution in a cluster of 
galaxies. We assume that there is a stationary homogeneous source of heat $q$ between the two Tribble 
boundary plates, which represent the cluster core and the cluster boundary. 
The heat is transported by electrons along the tangled magnetic field lines. We further assume that the thermal 
conductivity parallel to the field lines, $\kappa_\parallel$, is constant. 
We keep our assumption made in Section~\ref{BOUNDARY} that the cluster core is cold, i.e.~$T=0$ at ${\sf x}=1$. 
On the other hand, the density drops significantly at the cluster boundary, so we use the heat reflection 
condition there, $\partial T/\partial s=0$ at ${\sf x}=0$ (here, $s$ is the distance coordinate along a field line). 

Let consider a point $({\sf x}={\sf X},y,z)$ inside the cluster. The stationary temperature at this point 
depends on the lengths of the positive and the negative branches of the magnetic field line passing through 
this point. 
The positive branch always reaches the cluster core at ${\sf x}=1$, where it is cooled to zero temperature. 
Let the positive branch have dimensionless length ${\sf s}_1^+$. 
As for the negative branch, there are two possibilities. First, with probability 
${\tilde{\sf F}}_1^-(0)$ it reaches the core at ${\sf x}=1$ and cools down to $T=0$. In this case we 
denote the dimensionless length of the negative branch by ${\sf s}_1^-$. The second possibility is that the 
negative branch with probability ${\tilde{\sf F}}_0^-(0)$ reaches the cluster boundary at ${\sf x}=0$, where 
$\partial T/\partial {\sf s}=0$. In this case we denote the dimensionless length of the negative branch by 
${\sf s}_0^-$. In any case, the differential equation for the temperature distribution along the field line is
\beq
\frac{\partial^2 T}{\partial {\sf s}^2}+\frac{9X_0^4}{l_0^2}\frac{q}{\kappa_\parallel}=0,
\label{M_T_DIF_EQUATION}
\eeq
where ${\sf s}$ is the dimensionless coordinate along the field line, given by 
equations~(\ref{VARIABLES}).~\footnote{
Our description for the temperature distribution along the field lines really refers to a space average of 
temperature over a length interval long compared to the field decorrelation length, $l_0$. The effect of 
regions of weak and strong magnetic field (magnetic mirrors) is taken into account by introducing 
$\kappa_\parallel$, which is reduced relative to the Spitzer conductivity, see Malyshkin and Kulsrud 2000.
}

Now, let us consider a plane ${\sf x}={\sf X}$, which is perpendicular to the $x$-direction, and 
let us find the temperature averaged over points of this plane. We make this averaging by randomly choosing 
points of the plane, and then, by averaging temperature over the chosen 
points.~\footnote{
Formally, this plane is infinite, so are the number of points and distances between the points.
} 
The lengths ${\sf s}_1^+$, ${\sf s}_1^-$ and  ${\sf s}_0^-$ of positive and negative branches, going through 
different points are uncorrelated, and these lengths have the probability distributions ${\sf F}^+_1({\sf s})$, 
${\sf F}_1^-({\sf s})$ and ${\sf F}_0^-({\sf s})$ respectively. Therefore, the averaged temperature and the 
averaged temperature squared are
\beq
\langle T\rangle \!\!\!&=&\!\!\! \frac{q}{2\kappa_\parallel}\frac{9X_0^4}{l_0^2}\left\{
{\tilde{\sf F}}_1^-(0)\big\langle {\sf s}_1^+\big\rangle \big\langle {\sf s}_1^-\big\rangle + 
{\tilde{\sf F}}_0^-(0)\left[\big\langle ({\sf s}_1^+)^2\big\rangle + 
2\big\langle {\sf s}_1^+\big\rangle \big\langle {\sf s}_0^-\big\rangle\right]
\right\},
\label{M_T_AVERAGE}\\
\langle T^2\rangle \!\!\!&=&\!\!\! \left[\frac{q}{2\kappa_\parallel}\frac{9X_0^4}{l_0^2}\right]^2\!\left\{
{\tilde{\sf F}}_1^-(0) \big\langle({\sf s}_1^+)^2\big\rangle \big\langle({\sf s}_1^-)^2\big\rangle + 
{\tilde{\sf F}}_0^-(0)\left[\big\langle({\sf s}_1^+)^4\big\rangle + 
4\big\langle ({\sf s}_1^+)^3\big\rangle \big\langle {\sf s}_0^-\big\rangle + 
4\big\langle({\sf s}_1^+)^2\big\rangle \big\langle({\sf s}_0^-)^2\big\rangle\right]
\right\}.
\label{M_T2_AVERAGE}
\eeq
Here, the factors multiplying by ${\tilde{\sf F}}_1^-(0)$ represent the averaged temperature obtained by solving
equation~(\ref{M_T_DIF_EQUATION}) in the case when both the positive and the negative branches reach the cluster 
core (where $T=0$), as shown by the lower field line in Figure~\ref{FIG_PICTURE}. 
The factors multiplying by ${\tilde{\sf F}}_0^-(0)$ represent the averaged temperature 
obtained by solving equation~(\ref{M_T_DIF_EQUATION}) when the positive branch reaches the cluster core 
and the negative branch reaches the cluster boundary (where $\partial T/\partial {\sf s}=0$), as shown
by the upper field line in Figure~\ref{FIG_PICTURE}.
Remember that ${\tilde{\sf F}}_1^-(0)$ and ${\tilde{\sf F}}_0^-(0)$ are the probabilities for the negative 
branch to reach the core and the boundary respectively, while the positive branch always reaches the core.
All brackets $\langle ...\rangle$ mean averaging over appropriate probability distributions of lengths of 
branches. Note, that the expressions inside the brackets $\{...\}$ in the equations above are dimensionless, 
while $q$, $\kappa_\parallel$, $X_0$ and $l_0$ are not.

\begin{figure}
\vspace{7.0cm}
\includegraphics{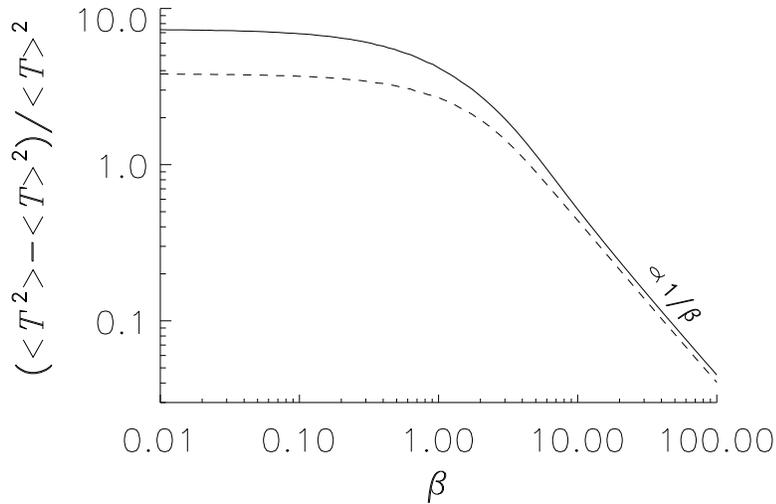}
\caption{The relative temperature dispersion, 
$(\langle T^2\rangle-\langle T\rangle^2)/\langle T\rangle^2$, versus $\beta$. 
The solid and the dashed lines represent $X=0.5X_0$ and $X=0.1X_0$ cases respectively.}
\label{FIG_DISPERSION}
\end{figure}

Now we use equations~(\ref{MOMENTS})--(\ref{FIRST_MOMENT_F_MINUS}) to express averaged powers of the branch 
lengths in the formulas~(\ref{M_T_AVERAGE}) and~(\ref{M_T2_AVERAGE}) in terms of the Laplace images of 
the probability distributions. We obtain
\beq
\langle T\rangle &=& \frac{q}{2\kappa_\parallel}\frac{9X_0^4}{l_0^2}\left\{
{\tilde{\sf F}}_0^-\frac{d^2{\tilde{\sf F}}_1^+}{d{\tilde{\sf s}}^2}+
\frac{d{\tilde{\sf F}}_1^-}{d{\tilde{\sf s}}}\frac{d{\tilde{\sf F}}_1^+}{d{\tilde{\sf s}}}+
2\frac{d{\tilde{\sf F}}_0^-}{d{\tilde{\sf s}}}\frac{d{\tilde{\sf F}}_1^+}{d{\tilde{\sf s}}}
\right\}_{{\tilde{\sf s}}=0},
\label{M_T}\\
\langle T^2\rangle &=& \left[\frac{q}{2\kappa_\parallel}\frac{9X_0^4}{l_0^2}\right]^2\left\{
{\tilde{\sf F}}_0^-\frac{d^4{\tilde{\sf F}}_1^+}{d{\tilde{\sf s}}^4}+
4\frac{d{\tilde{\sf F}}_0^-}{d{\tilde{\sf s}}}\frac{d^3{\tilde{\sf F}}_1^+}{d^3{\tilde{\sf s}}^3}+
\frac{d^2{\tilde{\sf F}}_1^-}{d{\tilde{\sf s}}^2}\frac{d^2{\tilde{\sf F}}_1^+}{d{\tilde{\sf s}}^2}+
4\frac{d^2{\tilde{\sf F}}_0^-}{d{\tilde{\sf s}}^2}\frac{d^2{\tilde{\sf F}}_1^+}{d{\tilde{\sf s}}^2}
\right\}_{{\tilde{\sf s}}=0}.
\label{M_T2}
\eeq
We can substitute the Laplace images of the three probability distributions given by 
formulas~(\ref{L_F_PLUS_1})--(\ref{L_F_MINUS_0}) into these equations and obtain analytical formulas
for $\langle T\rangle$ and $\langle T^2\rangle$. However, the resulting expressions are too complicated 
to be usefully interpreted. Therefore, we calculated all derivatives in the equations above numerically. 
Figure~\ref{FIG_DISPERSION} shows the relative dispersion of the temperature,
$(\langle T^2\rangle-\langle T\rangle^2)/\langle T\rangle^2$, as a function of parameter $\beta$ 
for two choices of $x$-position inside the cluster, $X=0.5X_0$ and $X=0.1X_0$.
We see that the dispersion is high for small values of $\beta$, while it is $\propto1/\beta\ll 1$ when
$\beta\gg 1$, according to the discussion in the last paragraph of Section~\ref{DIFFUSION_EQUATION}.

\begin{figure}
\vspace{6.7cm}
\includegraphics{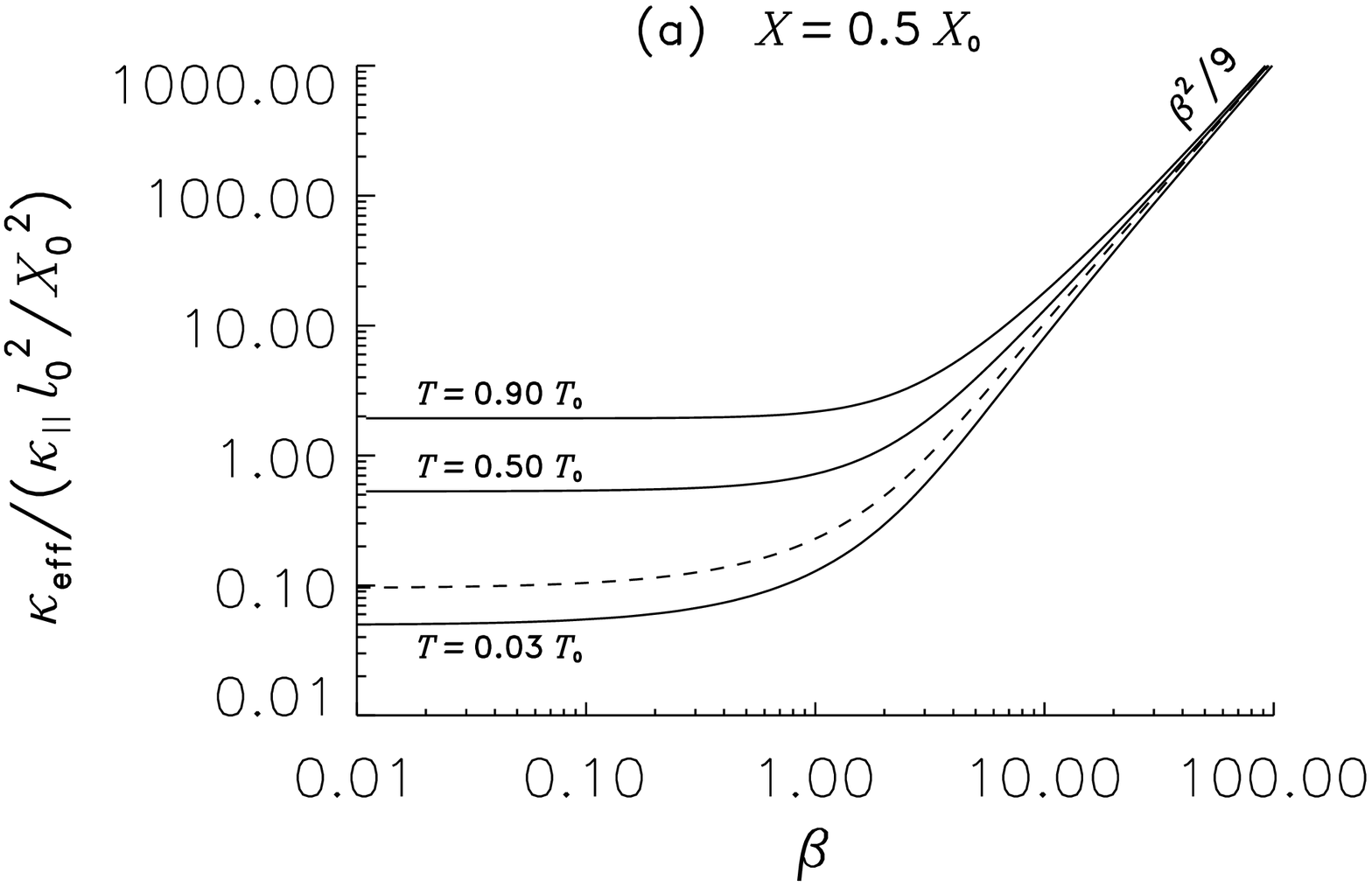}
\includegraphics{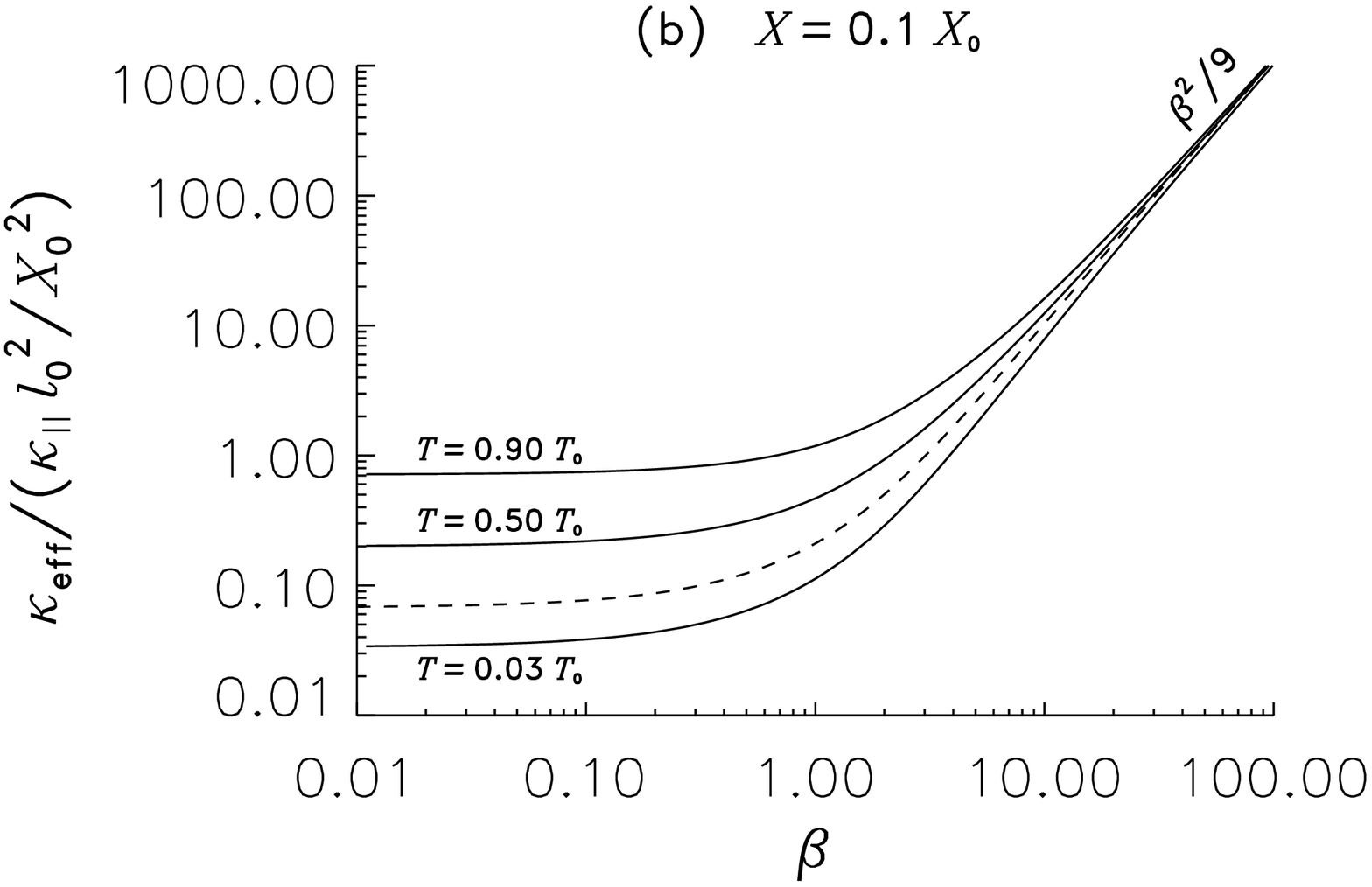}
\caption{The effective thermal conductivity $\kappa_{\rm eff}$ normalized to $\kappa_\parallel l_0^2/X_0^2$
is plotted as function of parameter $\beta$. The function $\kappa_{\rm eff}(\beta)$ is not universal and
it depends on how $\kappa_{\rm eff}$ is defined.}
\label{FIG_KAPPA_EFF}
\end{figure}

If there were no magnetic field at all, the temperature dispersion would be zero and the stationary 
temperature at ${\sf x}={\sf X}$ plane would be
\beq
T_{\rm S}=\frac{q}{2\kappa_{\rm eff}}(X_0^2-X^2),
\label{M_T_SPITZER}
\eeq
where the effective thermal conductivity $\kappa_{\rm eff}$ would be equal to the Spitzer thermal conductivity, 
$\kappa_{\rm eff}=\kappa_{\rm S}$.
Because there is tangled magnetic field, the actual stationary temperature is given by formula~(\ref{M_T}) and 
it is higher. However, instead of using formula~(\ref{M_T}) we can use the familiar formula~(\ref{M_T_SPITZER})
for the temperature if we choose the appropriate reduced value of the effective thermal conductivity.  
Equating formulas~(\ref{M_T}) and~(\ref{M_T_SPITZER}), we define the effective thermal conductivity for our
stationary model as
\beq
\kappa_{\rm eff}=
\kappa_\parallel \frac{l_0^2}{X_0^2}\,\frac{X_0^2-X^2}{9X_0^2}\left\{
{\tilde{\sf F}}_0^-\frac{d^2{\tilde{\sf F}}_1^+}{d{\tilde{\sf s}}^2}+
\frac{d{\tilde{\sf F}}_1^-}{d{\tilde{\sf s}}}\frac{d{\tilde{\sf F}}_1^+}{d{\tilde{\sf s}}}+
2\frac{d{\tilde{\sf F}}_0^-}{d{\tilde{\sf s}}}\frac{d{\tilde{\sf F}}_1^+}{d{\tilde{\sf s}}}
\right\}_{{\tilde{\sf s}}=0}^{-1}.
\label{M_KAPPA_EFF}
\eeq
The dashed lines in Figures~\ref{FIG_KAPPA_EFF}(a) and~(b) show this effective conductivity normalized
to $\kappa_\parallel l_0^2/X_0^2$ as a function of $\beta$ for the two choices of position inside the cluster, 
$X=0.1X_0$ and $X=0.5X_0$. To obtain the plots we again calculated all derivatives of the Laplace images
in equation~(\ref{M_KAPPA_EFF}) numerically. Note, that $\kappa_\parallel$ is the parallel thermal conductivity
reduced by random magnetic mirrors (Malyshkin and Kulsrud 2000).

Note, that for large values of $\beta$ the effective conductivity is independent of $X$ and is simply
$\kappa_{\rm eff}=\kappa_\parallel \beta^2l_0^2/9X_0^2=\kappa_\parallel(\Delta x/l_0)^2$.~\footnote{
This result can be found by the following calculations. The solutions of equations~(\ref{DIFFUSION_EQ}) 
in the limit $\beta\gg 1$ are ${\sf P}^\pm({\sf s},{\sf x})\approx \delta({\sf x}-{\sf X}\mp\beta {\sf s})$. 
Thus, all positive field line branches reach the core and have dimensionless lengths
${\sf s}_1^+\approx (1-{\sf X})/\beta$, while all negative branches reach the boundary and have 
lengths ${\sf s}_0^-\approx {\sf X}/\beta$. Substituting these expressions and 
${\tilde{\sf F}}_1^-(0)\approx 0$, ${\tilde{\sf F}}_0^-(0)\approx 1$ into formula~(\ref{M_T_AVERAGE}), and 
equating the result with formula~(\ref{M_T_SPITZER}), we obtain 
$\kappa_{\rm eff}=\kappa_\parallel \beta^2l_0^2/9X_0^2=\kappa_\parallel(\Delta x/l_0)^2$, see also 
eqs.~(\ref{BETA}) and~(\ref{DELTA_X}). This result is valid only if $\beta^2l_0^2/X_0^2\ll 1$, which is 
equivalent to the assumed condition $\epsilon=\langle B\rangle/\langle\delta B\rangle\ll 1$, see 
Sec.~\ref{DIFFUSION_EQUATION}.
}
This result exactly coincides with that given by the equation~(4) of Tao (1995), who gives this result
as a lower limit on $\kappa_{\rm eff}$.
On the other hand, for small and moderate values of $\beta$ the effective conductivity depends on $X$ as on
a parameter. This dependence results from our definition of $\kappa_{\rm eff}$ given above.
In other words, the actual value of $\kappa_{\rm eff}$ depends on how one defines it for a particular problem 
under consideration. The calculations in the next section further support this important statement.

%************************************************************************************************************
%************************************************************************************************************
%************************************************************************************************************

\section{Time evolution of temperature in clusters of galaxies}\label{T_EVOLUTION}

In the previous section we considered a stationary one-dimensional temperature distribution in a cluster of 
galaxies, assuming a constant heat source inside the cluster. In this section we solve a time-dependent problem 
and find the evolution of temperature of the cluster in time. 
Let assume, that when the cluster was formed at zero time, $t=0$, the temperature was homogeneous inside 
the cluster, i.e.~$T=T_0$, $0\le {\sf x}\le1$ (here ${\sf x}$ is the dimensionless 
$x$-coordinate inside the cluster; ${\sf x}=0$ and ${\sf x}=1$ correspond to the boundary and the core of 
the cluster respectively, see Figure~\ref{FIG_PICTURE}).
We assume that the cluster cools down in time by the heat conduction into the dense cluster core. 
As in the previous section, we again assume that the parallel thermal conductivity $\kappa_\parallel$ is 
constant, that the core is cold, $T=0$ at ${\sf x}=1$, and that there is the heat reflection condition 
along magnetic field lines at the cluster boundary, $\partial T/\partial {\sf s}=0$ at ${\sf x}=0$ 
(here ${\sf s}$ is the coordinate along field lines).

Because electrons travel along magnetic field lines, each field line cools down in time individually. 
Let us consider a temperature evolution at a point $({\sf x}={\sf X},y,z)$ inside the cluster. There is a 
single field line going through this point. Initially, at $t=0$, the temperature at this point is $T_0$. 
Then, as the field line cools down, the temperature drops in time. It is convenient to introduce a 
dimensionless time variable
\beq
\tau=\frac{l_0^2}{X_0^4}\frac{\kappa_\parallel}{\rho C_{\rm H}}\,t,
\label{TAU}
\eeq
where $\rho$ and $C_{\rm H}$ are the mass density and the heat capacity (per unit mass) of gas inside the 
cluster. For simplicity, we assume that the product $\rho C_{\rm H}$ is constant. Then, the time evolution 
of the temperature at the point $({\sf X},y,z)$ is given by the following simple equations:
\beq
\frac{\partial T(\tau,{\sf s})}{\partial \tau}=\frac{1}{9}\frac{\partial^2 T}{\partial^2 {\sf s}},
\qquad 
T(0,{\sf s})=T_0,
\label{T_EVOLUTION_EQ}
\eeq
where ${\sf s}$ is the dimensionless distance counted along the field line starting at the point 
$({\sf X},y,z)$, see equations~(\ref{VARIABLES}).

There are two possibilities for the field line going through the point $({\sf X},y,z)$. First, both the 
line's positive branch of length ${\sf s}_1^+$ and the line's negative branch of length ${\sf s}_1^-$ 
reach the cold cluster core (see the lower line in Figure~\ref{FIG_PICTURE}). 
In this case the boundary conditions on the temperature distribution along the line are
\beq
T(\tau,-{\sf s}_1^-)=0, \qquad T(\tau,{\sf s}_1^+)=0.
\nonumber
\eeq
With these boundary conditions the solution of equations~(\ref{T_EVOLUTION_EQ}) for the temperature at 
the point $({\sf X},y,z)$, where ${\sf s}=0$, is
\beq
T_{11}(\tau,{\sf s}_1^-,{\sf s}_1^+)=T_0\frac{4}{\pi}\sum\limits_{n=0}^\infty \, (1+2n)^{-1}
\exp{\left[-\frac{\pi^2\tau}{9({\sf s}_1^-+{\sf s}_1^+)^2}(1+2n)^2\right]}\,
\sin{\left[\frac{\pi {\sf s}_1^+}{{\sf s}_1^-+{\sf s}_1^+}(1+2n)\right]}.
\label{T_11}
\eeq
Here, the index ``$11$'' indicates that both branches reach the core at ${\sf x}=1$. The second possibility 
is that the line's negative branch of length ${\sf s}_0^-$ reaches the cluster boundary, while the line's 
positive branch of length ${\sf s}_1^+$ ends up in the cluster core (see the upper line in 
Figure~\ref{FIG_PICTURE}). In this case we have
\beq
(\partial T/\partial {\sf s})|_{{\sf s}\,=-{\sf s}_0^-}=0, \qquad T(\tau,{\sf s}_1^+)=0,
\nonumber
\eeq
and the solution of equations~(\ref{T_EVOLUTION_EQ}) for the temperature at the point $({\sf X},y,z)$, 
where ${\sf s}=0$, is
\beq
T_{01}(\tau,{\sf s}_0^-,{\sf s}_1^+)=T_0\frac{4}{\pi}\sum\limits_{n=0}^\infty \, (1+2n)^{-1}
\exp{\left[-\frac{\pi^2\tau}{36({\sf s}_0^-+{\sf s}_1^+)^2}(1+2n)^2\right]}\,
\sin{\left[\frac{\pi {\sf s}_1^+}{2({\sf s}_0^-+{\sf s}_1^+)}(1+2n)\right]}.
\label{T_01}
\eeq
Here, the index ``$01$'' indicates that the negative branch goes to the boundary at ${\sf x}=0$,
while the positive branch goes to the core at ${\sf x}=1$.

\begin{figure}
\vspace{6.7cm}
\includegraphics{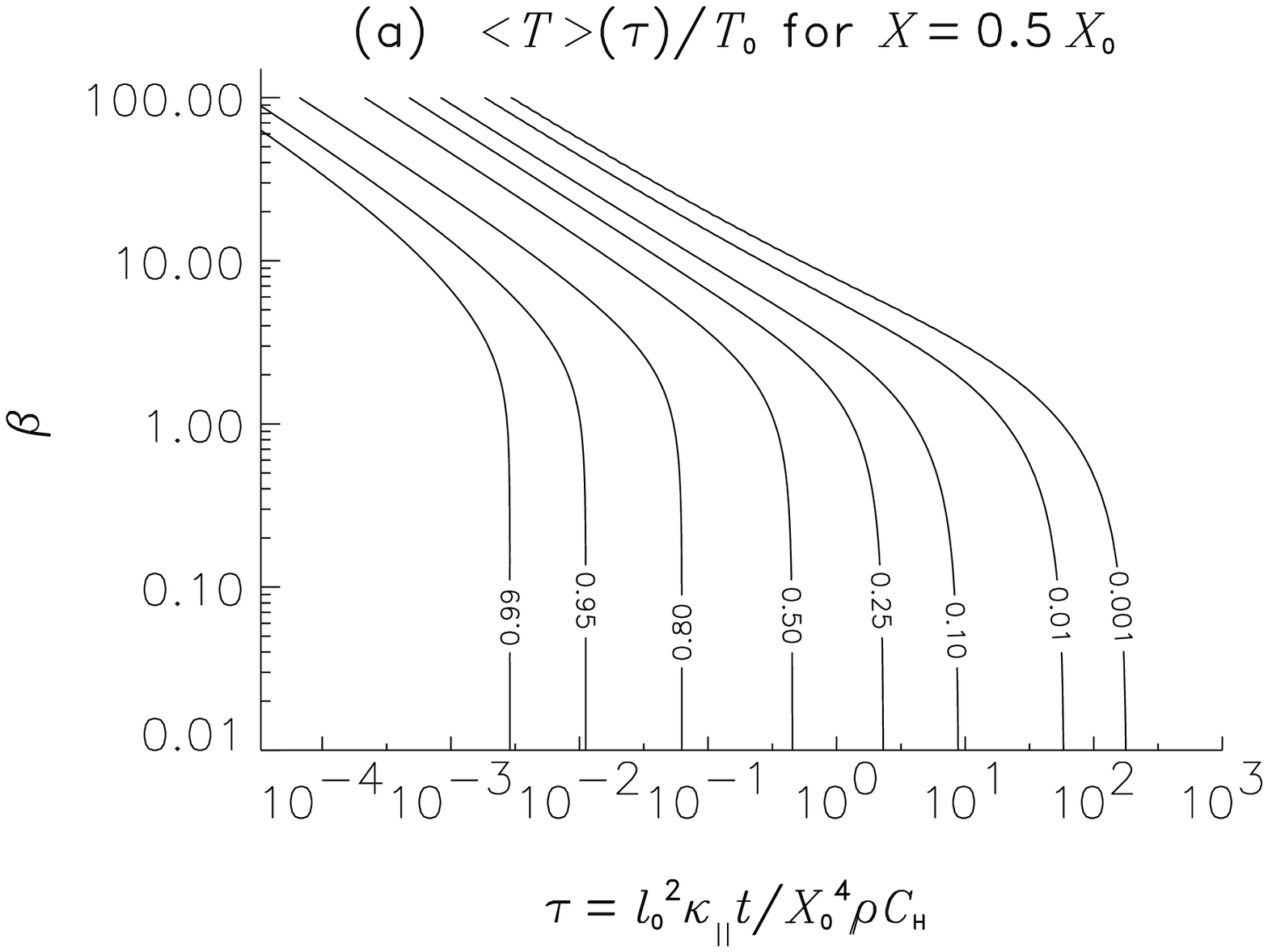}
\includegraphics{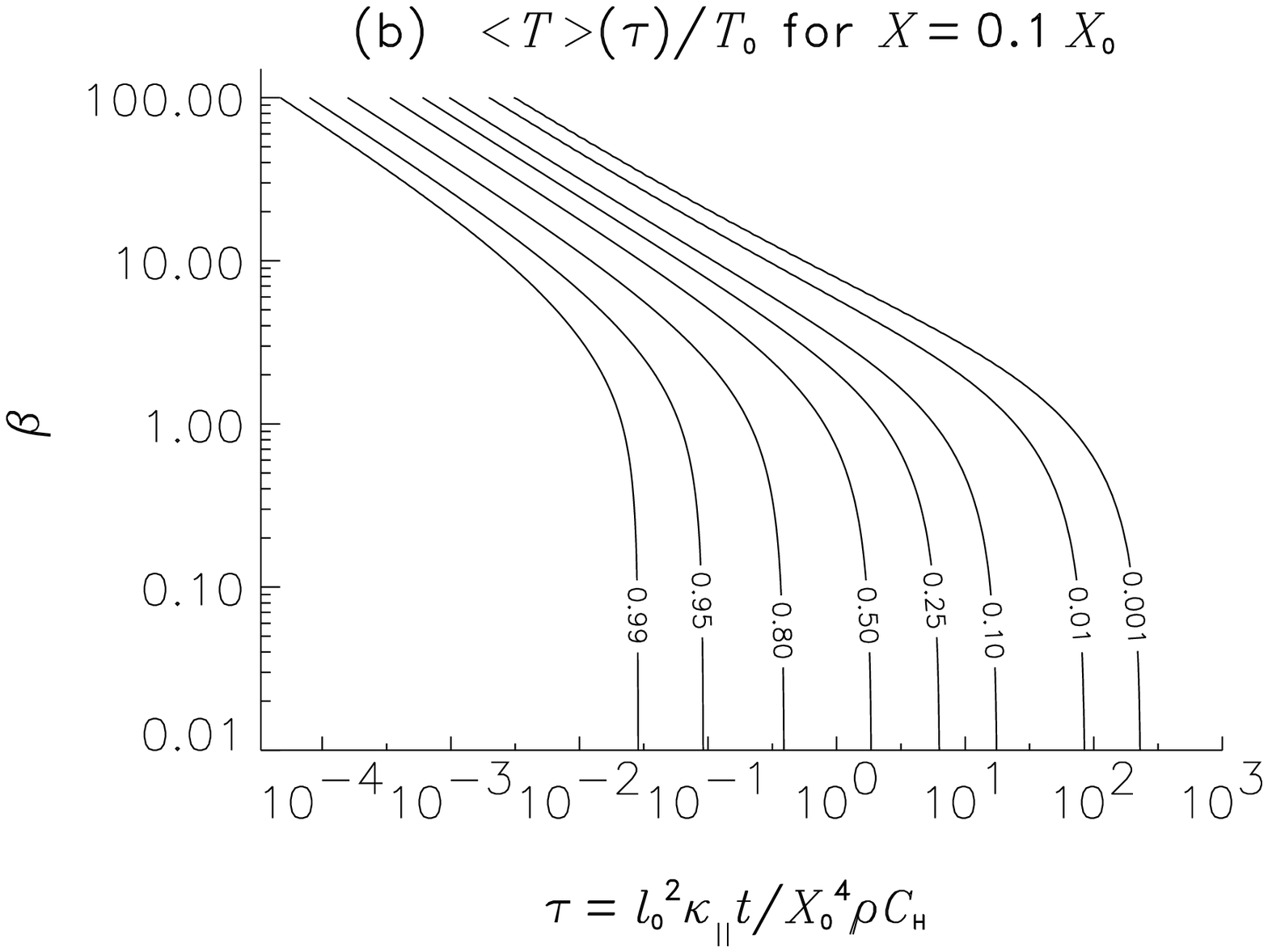}
\caption{Contour plots of the average temperature normalized to the initial temperature,
$\langle T\rangle(\tau)/T_0$, for two choices of $X$: (a) $X=0.5X_0$ and (b) $X=0.1X_0$. }
\label{FIG_T_MAP}
\end{figure}

Now, let find the temperature averaged over all points in the plane ${\sf x}={\sf X}$ in the same way
as we did in the previous section. We have
\beq
\langle T\rangle(\tau)=\int\limits_0^\infty\!\int\limits_0^\infty T_{11}(\tau,{\sf s}_1^-,{\sf s}_1^+)
{\sf F}_1^-({\sf s}_1^-) {\sf F}_1^+({\sf s}_1^+)\,d{\sf s}_1^-d{\sf s}_1^+ 
+\int\limits_0^\infty\!\int\limits_0^\infty T_{01}(\tau,{\sf s}_0^-,{\sf s}_1^+)
{\sf F}_0^-({\sf s}_0^-) {\sf F}_1^+({\sf s}_1^+)\,d{\sf s}_0^-d{\sf s}_1^+.
\label{T}
\eeq
Here, we have averaged over the probability distributions of ${\sf s}_1^+$, ${\sf s}_1^-$ and ${\sf s}_0^-$, 
which are given by equations~(\ref{F_PLUS}) and~(\ref{F_MINUS}). The two integrals in formula~(\ref{T}) 
correspond to the two possibilities for the field line branches, considered above. We have carried out
these integrals numerically.
Figures~\ref{FIG_T_MAP}(a) and~(b) show contour plots of $\langle T\rangle(\tau)/T_0$ in 
the $\tau$--$\beta$ parameter space for the two choices of $X$,
$X=0.5X_0$ and $X=0.1X_0$. To obtain these plots, first, we numerically
solved equations~(\ref{DIFFUSION_EQ}) with the boundary conditions~(\ref{BOUNDARY_MINUS}) 
and~(\ref{BOUNDARY_PLUS}), by making use of the standard implicit algorithm for partial differential 
equations.~\footnote{
We used a grid with equal steps $\Delta {\sf s}$ and $\Delta {\sf x}$ in ${\sf s}$- and ${\sf x}$-coordinates.
The goal was to find the tabulated solution of equation~(\ref{DIFFUSION_EQ}) at all grid points. The 
solution at $s=0$ is a delta-function. To find the solution at $s>0$, we used a method similar to mathematical 
induction. This method allowed us, after we obtained the tabulated solution at ${\sf s}$, to find the unknown 
tabulated solution at ${\sf s}+\Delta {\sf s}$ by the following algorithm. First, we wrote equation~(\ref{DIFFUSION_EQ}) 
as a finite difference operators, taking all ${\sf x}$ derivatives at ${\sf s}+\Delta {\sf s}$ (the implicit algorithm). 
As a result, we obtained a system of linear equations, which was a tridiagonal matrix. Second, we solved this system 
by the method of the Gaussian decomposition with backsubstitution, and found the solution at ${\sf s}+\Delta {\sf s}$.
}
Then we calculated the probability distributions~(\ref{F_PLUS}) and~(\ref{F_MINUS}) for different values of 
$\beta$. Finally, we substituted these distributions and formulas~(\ref{T_11}),~(\ref{T_01}) into 
equation~(\ref{T}), and we calculated the temperature for different values of $\tau$ and $\beta$.
(The Laplace transform method was not useful for this problem because we need more than simple moments of
the ${\sf s}_1^+$ etc.)

If there were no magnetic field, the time evolution of the temperature would be given by equations
\beq
\frac{\partial T}{\partial t}=\frac{\kappa_{\rm eff}}{\rho C_{\rm H}}\frac{\partial^2 T}{\partial^2 x},
\quad T(0,x)=T_0,
\quad (\partial T/\partial x)|_{x=0}=0, 
\quad T(t,X_0)=0,
%\label{}
\eeq
resulting in the following expression for the temperature similar to equation~(\ref{T_01}):
\beq
T_{\rm S}(\tau) &=& T_0\frac{4}{\pi}\sum\limits_{n=0}^\infty \, (1+2n)^{-1}
\exp{\left[\frac{\pi^2}{4}\frac{\kappa_{\rm eff} t}{X_0^2\rho C_{\rm H}}(1+2n)^2\right]}\,
\sin{\left[\frac{\pi(X_0-X)}{2X_0}(1+2n)\right]}
\nonumber \\
&=& T_0\frac{4}{\pi}\sum\limits_{n=0}^\infty \, (1+2n)^{-1}
\exp{\left[\frac{\pi^2 \tau}{4}\frac{X_0^2\kappa_{\rm eff}}{l_0^2\kappa_\parallel}(1+2n)^2\right]}\,
\sin{\left[\frac{\pi(X_0-X)}{2X_0}(1+2n)\right]},
\label{T_S}
\eeq
where we use definition~(\ref{TAU}). The effective thermal conductivity $\kappa_{\rm eff}$ would be equal to 
the Spitzer thermal conductivity, $\kappa_{\rm eff}=\kappa_{\rm S}$, if there were no magnetic field. 

For each set of values $T$, $\tau$, $\beta$ and $X$ there is a value of $\kappa_{\rm eff}$, which makes 
equations~(\ref{T}) and~(\ref{T_S}) agree. This value can be taken as the effective thermal conductivity, 
but it does depend on conditions.
In the case of the presence of the tangled magnetic field, instead of using the complicated formula~(\ref{T})
we can use the simple formula~(\ref{T_S}) with this $\kappa_{\rm eff}$ reduced appropriately from the Spitzer 
value. To find the reduced effective conductivity, we first choose fixed values of position $X$ and 
temperature $T$. Then, we find $\kappa_{\rm eff}$ as a function of $\beta$ in a such way, that both formula~(\ref{T}) 
and formula~(\ref{T_S}) give the same chosen value of temperature, $T$, at the same time $\tau$.
The calculation of $\kappa_{\rm eff}(\beta)$ was done numerically for two choices of $X$: $X=0.5X_0$ and $X=0.1X_0$;
and for three choices of $T$: $T=0.03\,T_0$, $T=0.50\,T_0$ and $T=0.90\,T_0$.
[In other words, we numerically found such functions $\kappa_{\rm eff}(\beta)$, that the temperature contours 
$T/T_0=0.03$, $T/T_0=0.50$ and $T/T_0=0.90$ appeared the same on Figures~\ref{FIG_T_MAP} when we used 
formulas~(\ref{T}) and~(\ref{T_S}) for the temperature.]
The resulting functions $\kappa_{\rm eff}(\beta)$ normalized to $\kappa_\parallel l_0^2/X_0^2$ are given by 
the solid lines in Figures~\ref{FIG_KAPPA_EFF}. We again see, that $\kappa_{\rm eff}(\beta)$ is not an universal 
function, and that it varies for different choices of $T$ and $X$, except when the $\beta$ is large. For smaller 
values of temperature the effective conductivity becomes less because very long field lines, which keep the initial 
temperature, become more important.

%************************************************************************************************************
%************************************************************************************************************
%************************************************************************************************************

\section{Discussion}\label{DISCUSSION}

We would like to start the discussion with by stressing one of the main results we found: the actual value of 
this effective conductivity in the tangled magnetic fields depends on how this effective conductivity is defined. 
In our paper we used two ``natural'' definitions of $\kappa_{\rm eff}$ for two very simple models of a 
cluster of galaxies: a time-independent model in Section~\ref{MODEL} and a time-dependent model in 
Section~\ref{T_EVOLUTION}. The results for $\kappa_{\rm eff}$, reported in figures~\ref{FIG_KAPPA_EFF}(a)
and~\ref{FIG_KAPPA_EFF}(b), are the same only provided $\beta\gg 1$, and they are significantly different 
(by up to a factor of ten) when $\beta\simlt 1$. We conclude that there does not exist an universal result 
for the effective conductivity in the tangled magnetic fields. One has to define $\kappa_{\rm eff}$ and to 
calculate it for a particular astrophysical problem that he/she considers. We believe that such calculation 
is possible for many problems, including numerical simulations of galaxy cluster formation and of cooling 
flows, by making use of our diffusion approximation method for the random walk of tangled magnetic field lines.

It is useful to compare our results for the effective conductivity in the tangled magnetic fields with those
reported in previous papers. As we said in the last paragraph of Section~\ref{MODEL}, in the limit $\beta\gg 1$
our result for $\kappa_{\rm eff}$ coincides with the result obtained by Tao (1995). 
This simple formula (for $\beta$ large) corresponds to lines being in order $X_0/\epsilon$ in length, a result
sometimes believed (remember, that $\epsilon\ll 1$ is the ration of the field mean component to the field random 
component). But in fact, it is only valid as a lower limit on $\kappa_{\rm eff}$, except for large $\beta$.
However, we can speculate that it revolves around the very definition of $\epsilon$. We give $\epsilon$ first
and then impose the statistics of the field. If the statistics is given first and the mean field is defined
afterwards, as the rms value of the mean field on the scales $\sim X_0$, then $\kappa_{\rm eff}$ would be 
closer to our results.
 
In the limit $\beta\ll 1$, our result, $\kappa_{\rm eff}\sim 0.1 \kappa_{\parallel} l_0^2/X_0^2$, 
is consistent with the result of Tribble (1989).~\footnote{
Tribble has $\kappa_{\rm eff}=4\kappa_{\parallel} l_0^2/X_0^2$, see his equation~(11). The difference arises 
because: first, our boundary conditions~(\ref{BOUNDARY_MINUS}) and~(\ref{BOUNDARY_PLUS}) for the random walk 
of field lines are different from those of Tribble; second, the magnetic field lines in our model are allowed 
to random walk in three dimensions, while Tribble considered a restricted one-dimensional random walk.
}
To the best of our knowledge, there have no results for $\kappa_{\rm eff}$ when $\beta\sim 1$ obtained before.

Recently Chandran and Cowley (1998) have suggested that the diffusion of the heat conducting electrons 
perpendicular to the magnetic field lines is crucially important for the heat transport. In their model the 
statistically independent random step by the electrons is equal to $\sqrt{2L_{\rm RR}l_0}$, where 
$L_{\rm RR}\sim l_0\ln{(l_0/\rho_e)}$ is the Rechester-Rosenbluth length. This length is the distance along 
magnetic field lines over which a separation between two initially neighboring lines grows exponentially from 
the electron gyro-radius $\rho_e$ up to the field decorrelation length $l_0$ (because of the Kolmogorov-Lyapunov 
exponential divergence of field lines). 
Chandran and Cowley suggested that the perpendicular diffusion of electrons and this exponential divergence of 
the magnetic field lines enable the electrons to conduct heat between different field lines, which may 
enhance the effective thermal conductivity considerably. However, we believe that the picture considered by 
Chandran and Cowley is rigorously valid only if electrons can move a distance $L_{\rm RR}$ along field 
lines without collisions. In fact, there are collisions, and if the electron mean free path $\lambda$ is less 
than the Rechester-Rosenbluth length $L_{\rm RR}$, the exponential divergence of field lines does not help
electrons to diffuse perpendicular field lines. In clusters of galaxies, the ratio of $L_{\rm RR}$ to $\lambda$ 
is $L_{\rm RR}/\lambda \approx 5000\,(l_0/10\,{\rm kpc})\,(T/10^7{\rm K})^{-2}\,(n/10^{-3}{\rm cm}^{-3})$.
As a result, the perpendicular heat transport can generally be neglected (we accept $3$--$10\,{\rm kpc}$ as
a typical scale for magnetic fields in clusters of galaxies, see Kronberg 1994, Eilek 1999).
The Chandran and Cowley's model can be very important for some other astrophysical problems, but the
consideration of their model is beyond of the scope of this paper.

Finally, let estimate the importance of the heat conduction for the formation and evolution of galaxy 
clusters, using figures~\ref{FIG_T_MAP}. The parameter $\beta$ and the dimensional time $\tau$, given by 
equations~(\ref{BETA}) and~(\ref{TAU}), can be estimated as
\beq
\beta=20\,\frac{X_0}{1\,{\rm Mpc}}\,\frac{10\,{\rm kpc}}{l_0}\,\frac{\epsilon\cos\theta}{0.1},
\quad
\tau=5\!\times\!10^{-6}\,\frac{\kappa_\parallel}{\kappa_{\rm S}}\,\frac{t}{10^{10}{\rm yrs}}\,
\frac{10^{-3}{\rm cm}^3}{n}\left[\frac{T}{10^7K}\right]^{5/2}
\left[\frac{l_0}{10\,{\rm kpc}}\right]^2\left[\frac{1\,{\rm Mpc}}{X_0}\right]^4.
\nonumber
\eeq
The reduction of the parallel thermal conductivity relative to the Spitzer value, caused by magnetic mirrors,
depends on the ratio $l_0/\lambda=160\,(l_0/10\,{\rm kpc})\,(T/10^7{\rm K})^{-2}\,(n/10^{-3}{\rm cm}^{-3})$
[Malyshkin \& Kulsrud 2000]. The typical values are $\kappa_\parallel/\kappa_{\rm S}\sim 1/10$ for the hot and 
low-density halo of a galaxy cluster, and $\kappa_\parallel/\kappa_{\rm S}\sim 1$ for the cluster central 
region. We see that for the halo, where $X_0\sim 1\,{\rm Mpc}$, $l_0\sim 10\,{\rm kpc}$, 
$T\sim 5\times 10^7\,{\rm K}$ and $n\sim 2\times 10^{-4}{\rm cm}^{-3}$, we have $\beta\sim 20$ and 
$\tau\sim 10^{-4}$. Thus, according to figures~\ref{FIG_T_MAP}, the heat conduction is unimportant 
there, while according to figures~\ref{FIG_KAPPA_EFF}, the effective thermal conductivity is 
reduced by a factor of $\sim 2000$. At the same time, for the cluster central region we have 
$X_0\sim 0.1\,{\rm Mpc}$, $l_0\sim 3\,{\rm kpc}$, $T\sim 2\times 10^7\,{\rm K}$, 
$n\sim 10^{-3}\,{\rm cm}^{-3}$, $\beta\sim 7$ and $\tau\sim 0.03$, so the heat conduction is 
important there, and the effective thermal conductivity is reduced by a factor of $\sim 200$. This 
conclusion agrees with those of Tau (1995) and of Rosner and Tucker (1989). Thus, with certain parameters,
such as these, thermal conduction can be important even with such a large reduction factor.
Figures~\ref{FIG_T_MAP}(a) and~\ref{FIG_T_MAP}(b) give the direct temperature evolution without reference
to reduction factors, and perhaps because of the sensitivity of the reduction factor to its definition,
they are probably the more useful expressions of thermal evolution.

Suginohara and Ostriker in their hydrodynamic simulations of galaxy cluster formation encountered 
a  ``cooling catastrophe'', which appears as a steep non-realistic rise in the density profile of the 
relaxed core of a galaxy cluster because of the fast cooling in the core (Suginohara \& Ostriker 1998).
The heat conduction into the core was among solutions suggested by them. However, they believed that the
conduction can be neglected if it is reduced by a factor of $30$ or more. In this paper we find that 
the thermal conduction is very important in the cluster central region, despite the fact that in our models 
it is reduced by a factor of $\sim 200$. Thus, the conduction should be included in hydrodynamic 
simulations of galaxy clusters, even when the reduction factor is large.

\acknowledgments

I would like to especially thank my advisor, Professor Russell Kulsrud, for suggesting this problem,
for many interesting and very fruitful discussions of it, and for reading carefully this manuscript. 
I would also like to thank Professors Jeremiah Ostriker, Jeremy Goodman and David Spergel for their 
insightful comments. I am also very grateful to Professor Bruce Drain for financial support under 
NSF grants AST-9619429 and AST-9988126.

%************************************************************************************************************
%************************************************************************************************************
%************************************************************************************************************

{}

%************************************************************************************************************

\end{document}